\documentclass[final,5p,times,twocolumn]{elsarticle}

\usepackage[T1]{fontenc}
\usepackage[utf8]{inputenc}

\usepackage{amsmath}
\usepackage{amsfonts}
\usepackage{amssymb}
\usepackage{amsxtra}
\usepackage{array}
\usepackage{color}
\usepackage{dcolumn}
\usepackage{enumerate}
\usepackage{graphicx}
\usepackage{hepunits}
\usepackage{xspace}
\usepackage{CJKutf8}
\usepackage{scalerel}

\definecolor{purple}{rgb}{0.5,0,0.5}
\definecolor{blue}{rgb}{0.0,0,0.9}
\definecolor{prdblue}{rgb}{0.133,0.118,0.498}
\usepackage[colorlinks=true, pdfstartview=FitV, linkcolor=prdblue, citecolor= prdblue, urlcolor=prdblue]{hyperref}

\usepackage[mathscr,scaled=1.15]{urwchancal}
\DeclareFontFamily{OT1}{pzc}{}
\DeclareFontShape{OT1}{pzc}{m}{it}%
{<-> s * [1.15] pzcmi7t}{}
\DeclareMathAlphabet{\mathpzc}{OT1}{pzc}{m}{it}

\def\srm#1{{\rm{\scriptscriptstyle #1}}}

\newcommand{\be}{\begin{equation}}
\newcommand{\bea}{\begin{eqnarray}}
\newcommand{\ee}{\end{equation}}
\newcommand{\eea}{\end{eqnarray}}

\def\1eq#1{Eq.~(\ref{#1})}

\def\2eqs#1#2{Eqs.~(\ref{#1}) and~(\ref{#2})}
\def\3eqs#1#2#3{Eqs.~(\ref{#1}),~(\ref{#2}) and~(\ref{#3})}

\def\fig#1{Fig.~\ref{#1}}

\def\ie{{\it i.e.}, }
\def\eg{{\it e.g.}, }

\def\s#1{{\scriptscriptstyle #1}}

\newcommand{\fatg}{{\rm{I}}\!\Gamma}

\newcommand\wh[1]{\hstretch{2}{\hat{\hstretch{.5}{#1}}}}

\biboptions{sort&compress}

\journal{Physics Letters B}

\begin{document}
\begin{CJK}{UTF8}{song}

\begin{frontmatter}


%
\title{Lattice determination of the Batalin-Vilkovisky function \\ and the strong running interaction}

\author[Campinas]{A.~C.~Aguilar}

\author[Coimbra,Plymouth]{N.~Brito}

\author[UV]{M.~N.~Ferreira}

\author[UV]{J. ~Papavassiliou}

\author[Coimbra]{O.~ Oliveira}

\author[Coimbra]{P.~J.~ Silva}

\address[Campinas]{
University of Campinas - UNICAMP, Institute of Physics Gleb Wataghin, 13083-859 Campinas, S\~ao Paulo, Brazil}

\address[Coimbra]{
CFisUC,  Department of Physics, University of Coimbra, Coimbra, 3004-516, Portugal}

\address[Plymouth]{Centre for Mathematical Sciences, University of Plymouth, Plymouth, PL4 8AA, United Kingdom}

\address[UV]{Department of Theoretical Physics and IFIC, University of Valencia and CSIC, E-46100, Valencia, Spain}

%

%
%


\begin{abstract}

The Batalin-Vilkovisky function
is a central component in 
the modern formulation 
of the background field method  
and the 
physical applications derived 
from it.
In the present work 
we report on novel 
lattice results for this particular quantity, obtained 
by capitalizing on 
its equality with 
the Kugo-Ojima function in the Landau gauge. 
The results 
of the lattice simulation are in very good agreement with the predictions derived  
from a continuum analysis based on the 
corresponding Schwinger-Dyson equations.
In addition, we show that an important 
relation connecting this function 
with the ghost propagator is fulfilled rather accurately.
With the aid of these results, 
we carry out the first completely 
lattice-based determination of the 
process-independent strong running interaction, employed in a variety of 
phenomenological studies. 

\end{abstract}

\begin{keyword}
 Lattice QCD \sep Background Field Method \sep
Schwinger-Dyson equations \sep
strong running interaction  \sep

\end{keyword}

\end{frontmatter}
\end{CJK}

\section{Introduction}
\label{introduction}

The background field method (BFM)
is a powerful framework that permits the 
implementation of  the gauge-fixing procedure necessary for
quantizing gauge theories without losing explicit gauge invariance~\cite{DeWitt:1967ub,
Honerkamp:1972fd,Kallosh:1974yh,Kluberg-Stern:1974nmx,Arefeva:1974jv,
Abbott:1980hw,Weinberg:1980wa,Abbott:1981ke,Shore:1981mj,Abbott:1983zw}.
The formulation of non-Abelian gauge theories 
within this quantization scheme affords a plethora of advantages, 
both from the formal as well as the practical points of view. 
Thus, in addition to streamlining a variety of demonstrations related 
to renormalization, 
it allows one to tackle efficiently longstanding challenges,  
such as the gauge-invariant truncation 
of the Schwinger-Dyson equations (SDEs)
\cite{Roberts:1994dr,Alkofer:2000wg,Fischer:2006ub,Roberts:2007ji,Binosi:2009qm,Cloet:2013jya,
Huber:2018ned}
or the definition of 
physically meaningful renormalization-group invariant (RGI) 
quantities, see, \eg~\cite{Binosi:2009qm}.   

Intrinsic to the BFM formalism is the   
duplication of the 
number of gauge fields,
which are distinguished into 
``background'' and "quantum" types
~\cite{Abbott:1981ke,Abbott:1983zw}; while the former are not integrated 
over in the path integral and do not appear in loops, 
the latter are identified with the 
standard gauge fields known from the 
conventional formulation, \eg in the linear 
covariant gauges~\cite{Fujikawa:1972fe}.

The correlation (Green's) functions built out of  
background fields 
satisfy linear 
Slavnov-Taylor identities (STIs)~\cite{Taylor:1971ff,Slavnov:1972fg},  which are naive generalizations of 
tree level  relations, without  deformations originating from the 
ghost-sector of the theory.
In addition, and more importantly 
for our purposes, they 
are connected to the corresponding 
Green's functions  
involving quantum fields by 
exact relations, known as "background-quantum 
identities" (BQIs)~\cite{Grassi:1999tp,Grassi:2001zz,Binosi:2002ez,Binosi:2009qm}. These special identities are  
derived with the aid of the Batalin-Vilkovisky 
(BV) formalism~\cite{Batalin:1977pb,Batalin:1983ggl}, where  
the original BFM action is extended through the inclusion of 
suitable anti-fields and sources.  

The common ingredient of this infinite tower of identities is a particular function, denoted by $G(q)$,
which we 
denominate "BV function"~\cite{Binosi:2002ez}. The precise field-theoretic definition 
of this function involves a special combination of an 
anti-field and a source; however, a more practical 
relation expresses $G(q)$ in terms of the conventional 
gluon and ghost propagators, and the ghost-gluon kernel,
known from the STI satisfied by the 
three-gluon vertex~\cite{Marciano:1977su,Ball:1980ax,Pascual:1984zb,Davydychev:1996pb,Gracey:2011vw,Gracey:2014mpa, Aguilar:2018csq,Aguilar:2019jsj}.

The BV function appears in a variety of non-perturbative applications,
such as the so-called "block-wise" truncations of the SDE series~\cite{Aguilar:2006gr,Binosi:2007pi,Binosi:2008qk,Aguilar:2008xm,Aguilar:2022exk}, 
and the definition of an effective interaction which is both 
RGI and process-independent~\cite{Binosi:2002vk,Aguilar:2009nf}, constituting a common component of 
every two-to-two on-shell process. In addition, $G(q)$ is related to 
the inverse of the ghost dressing function through an exact 
relation~\cite{Kugo:1995km,Grassi:2004yq,Kondo:2009ug,Aguilar:2009nf}, 
which enforces the coincidence of distinct versions of 
effective charges 
in the deep infrared~\cite{Aguilar:2009nf}.
In that sense, the BV function represents an essential  
component of the gauge sector of Yang-Mills theories, participating non-trivially in the description of the  
associated dynamics. 

Particularly pivotal to the present  analysis is the {\it Landau-gauge}
equality 
between the BV function and the so-called Kugo-Ojima (KO) 
function~\cite{Kugo:1979gm,Kugo:1995km}, whose field-theoretic definition is especially suitable for a lattice simulation~\cite{Nakajima:1999dq,Sternbeck:2006rd,Brito:2023pqm,Brito:2023rfv}. 
Exploiting this key relation, we carry out a large-volume 
simulation of the KO function, and thus,
we simultaneously obtain the lattice result for the 
BV function. 

The lattice results are contrasted with those obtained from the corresponding 
SDE that governs the evolution of $G(q)$, finding rather good agreement. In addition, the aforementioned 
relation between $G(q)$ and the ghost dressing function is shown to 
be fulfilled 
at a reasonable level of accuracy. 
Furthermore, the lattice results for 
$G$, together with the 
data for the gluon propagator obtained from the same lattice, 
allow us to 
present the first purely lattice-based 
construction of the effective interaction 
introduced in~\cite{Binosi:2014aea,Cui:2019dwv}.

\section{The ubiquitous Batalin-Vilkovisky function }\label{sec:BVBFM}

Within the BFM approach,  
the gauge field $A$ appearing in the classical action is decomposed as $A = B +Q$, where $B$ and $Q$ are the 
background and quantum (fluctuating) fields, respectively. 
 Background fields 
participate in Feynman diagrams only as external legs, while loops  
are comprised exclusively by quantum fields~\cite{Abbott:1981ke,Abbott:1983zw}. 
The key property of 
the BFM 
is that the gauge-fixing may be implemented without compromising 
explicit gauge invariance. 
Specifically, 
instead of the gauge-fixing term  
\be 
{\cal L}_{\mathrm{gf}} = \frac{1}{2\xi} (\partial_\mu Q^{a \mu})^2 \,,
\label{Rgf}
\ee
of the  conventional 
covariant ($R_\xi$)  gauges~\cite{Fujikawa:1972fe}, one uses  
\be 
{\widehat{\cal L}}_{\mathrm{gf}} = \frac{1}{2\xi_{Q}}({\wh D}_\mu^{ab}Q^{b \mu})^2 \,, \qquad {\wh D}_\mu^{ab} = \partial_\mu \delta^{ab} + g f^{amb} B_\mu^m \,,
\label{gfBFM}
\ee 
where $f^{abc}$ are the structure constants of SU$(3)$. 
The key feature that converts the BFM into a powerful quantization scheme is that the gauge-fixing choice of \1eq{gfBFM}
gives rise to a gauge-fixed action which is 
invariant under gauge transformations of the background field, 
$\delta B^{a}_{\mu} = -g^{-1} \partial_{\mu} \theta^{a} 
+ f^{abc} \theta^{b} B^{c}_{\mu}$, 
where $g$ denotes the gauge coupling. 

A profound consequence of this invariance is the 
form of the resulting STIs. Specifically, 
when Green's functions are contracted 
by the momentum carried by one of their background gluons,  
they satisfy {\it ghost-free} STIs, 
akin to the Takahashi identities~\cite{Ward:1950xp,Takahashi:1957xn} known from Abelian theories, such as QED. 
These special STIs have far-reaching consequences 
for renormalization, because they impose 
QED-like constraints among the 
various renormalization constants.
In particular, 
the wave-function renormalization 
constant of the $B$ field, ${Z}_B$,
and the gauge coupling renormalization constant, 
$Z_g$, defined as 
\be
B^{\,a \mu}_{{\rm \s{R}}} = {Z}_B^{-1/2} B^{\,a\mu} \,,
\qquad \qquad g_{{\rm \s{R}}} = Z_g^{-1} g \,,
\label{Zg_ZA_def}
\ee
satisfy the key relation (see, \eg~\cite{Abbott:1981ke,Abbott:1983zw})
\be 
Z_g = {Z}_B^{-1/2} \,,
\label{Zg_ZA}
\ee
in close analogy to the textbook relation 
$Z_e = Z_A^{-1/2}$, obeyed by 
the renormalization constants of the electric charge 
and the photon~\cite{Itzykson:1980rh}. 

In what follows we will identify the quantum gauge-fixing parameter 
$\xi_Q$ of the BFM [see \1eq{gfBFM}] 
with the corresponding  parameter $\xi$ 
of the covariant gauges [see \1eq{Rgf}], 
\ie $\xi_Q =\xi$; 
in particular, in the Landau
gauge that we adopt throughout this work, $\xi_Q =\xi=0$.
With this identification, 
the propagator $\Delta^{ab}_{\mu\nu}(q)=-i\delta^{ab}\Delta_{\mu\nu}(q)$
that connects two quantum gluons coincides with the standard gluon propagator 
of the $R_{\xi}$ gauges; in the Landau gauge, $\Delta_{\mu\nu}(q)$ 
is completely transverse, \ie 
\be
\Delta_{\mu\nu}(q) = \Delta(q) {P}_{\mu\nu}(q)\,, \qquad {P}_{\mu\nu}(q) := g_{\mu\nu} - q_\mu q_\nu/{q^2}\,.
\label{defgl}
\ee
In addition to $\Delta^{ab}_{\mu\nu}(q)$, in the BFM we have 
two more gluon propagators, one connecting 
the fields $Q^a_\mu(q)$ with a $B^b_\nu(-q)$, and 
one connecting $B^a_\mu(q)$ with a $B^b_\nu(-q)$; the latter is essential 
for the ensuing analysis, and  
will be denoted by
\be
\wh{\Delta}^{ab}_{\mu\nu}(q)=-i\delta^{ab}\wh{\Delta}(q) P_{\mu\nu}(q).
\label{deltahat}
\ee

\begin{figure}[t]
  \centering
  \includegraphics[width=0.4\textwidth]{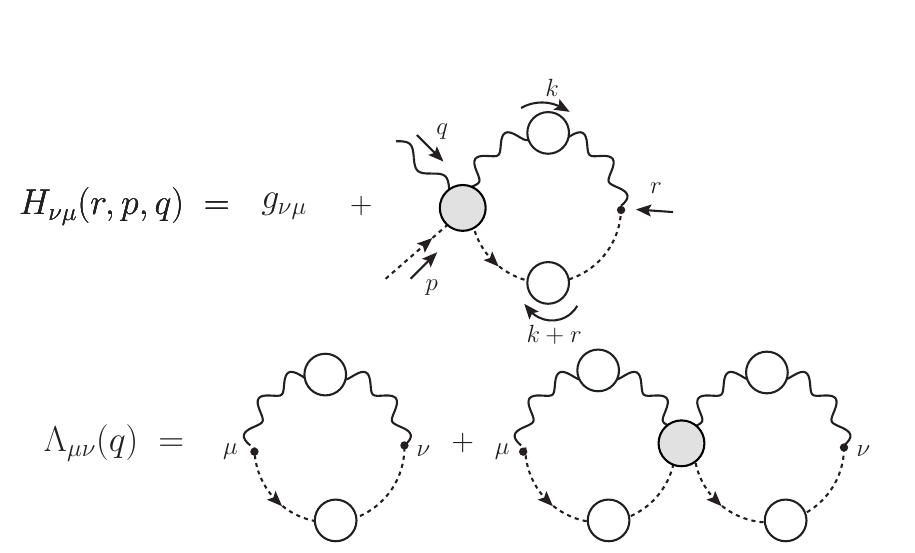}
\caption{ Diagrammatic definitions of the ghost-gluon scattering kernel, $H_{\nu\mu}(r,p,q)$, and the two-point function $\Lambda_{\mu\nu}(q)$, given in Eq.~\eqref{Lambda_GL}.  }
\label{fig:H_def}
\end{figure}

It turns out that the gluon propagators $\Delta(q)$ and $\wh\Delta(q)$
are related by the simplest of all BQIs, namely~\cite{Grassi:1999tp,Binosi:2002ez,Grassi:2004yq,Binosi:2013cea} 
\be
\Delta(q) = [1 + G(q)]^2 \, \wh\Delta(q)\,. 
\label{propBQI}
\ee
$G(q)$ is the $g_{\mu\nu}$ component of a certain  two-point function, $\Lambda_{\mu\nu}(q)$,
 given by (Minkowski space)
\bea 
\Lambda_{\mu\nu}(q) &:=& i g^2 C_{\rm A}\int_k \Delta^\rho_\mu(k)D(k+q)H_{\nu\rho}(-q,k+q,-k) 
\nonumber\\
&=& \underbrace{G(q)}_{\rm BV \,function} \!\!\!\!g_{\mu\nu}+ \,L(q) \,\frac{q_\mu q_\nu}{q^2} \,, \label{Lambda_GL}
\eea 
where $C_\mathrm{A}$ is the Casimir eigenvalue of the adjoint representation [$N$ for SU$(N)$], 
$D^{ab}(q) = 
i \delta^{ab} D(q)$ is the 
ghost propagator, 
and $H_{\nu\mu}(r,p,q)$ denotes the ghost-gluon kernel, see upper panel of \fig{fig:H_def}.
Note that, 
formally, 
$\Lambda_{\mu\nu}(q)$  
is a two-point function  
of a background source, 
$\Omega_{\mu}^{a}$, and a 
gluon anti-field, $A_{\nu}^{b\star}$,
usually denoted in the literature by 
$\fatg_{\Omega_{\mu}^{a} A_{\nu}^{b\star}}(q)$; 
the version given in \1eq{Lambda_GL} 
[see lower panel of \fig{fig:H_def}]
relies on the fact that 
$\Omega_{\mu}^{a}$ and 
$A_{\nu}^{b\star}$ may be replaced
by the  Becchi-Rouet-Stora-Tyutin (BRST)~\cite{Becchi:1975nq,Tyutin:1975qk} composite operator to which they are coupled.
 
Next, consider three-point functions 
(vertices) of the general type 
${\mathbf \Gamma}_{B X Y }$ and 
${\mathbf \Gamma}_{A X Y }$ 
where $X $ and $Y$ represent general fields 
(\eg $X=Y=A$ in the case of a three-gluon vertex,
or $X=\psi$, $Y = {\bar \psi}$ for 
the quark-gluon vertex). The vertex BQI  
has the general form
\be
{\mathbf \Gamma}_{B X Y } = [ 1 + G(q) ] \, {\mathbf \Gamma}_{A X Y }
+ \cdots \,, 
\label{quark_BQI}
\ee
where the ellipsis denotes terms 
that involve additional auxiliary functions, 
and vanish "on shell". 

Higher correlation functions 
are related by similar, albeit increasingly more complicated BQIs,  
with the BV function playing 
always a prominent role~\cite{Binosi:2008qk}.  

We next turn to a special identity 
that relates 
the two form factors 
$G(q)$ and $L(q)$ of 
$\Lambda_{\mu\nu}(q)$  
with the ghost dressing function, 
$F(q)$, defined as  
\be
F(q) = q^2 D(q) \,.
\label{theF}
\ee

Specifically,
in the Landau gauge, 
we have the exact relation 
~\cite{Aguilar:2009nf,Binosi:2013cea,Binosi:2014aea}
\be 
F^{-1}(q) = 1 + G(q) + L(q) \,. \label{FGL}
\ee 

Due to the fact that 
the function $L(q)$ vanishes 
at the origin, $L(0) =0$~\cite{Aguilar:2009nf}, 
from \1eq{FGL} we obtain the important result
\be 
F^{-1}(0) = 1 + G(0)\,.
\label{FG0}
\ee
Note that this last relation enforces the 
coincidence 
between the Taylor
~\cite{vonSmekal:1997ohs,Alkofer:2000wg,Fischer:2003rp,Bloch:2003sk,Oliveira:2006zg,Sternbeck:2007br,Boucaud:2008gn,Horak:2023xfb,Cyrol:2017ewj,Corell:2018yil,Deur:2023dzc,deTeramond:2024ikl} and 
"pinch-technique"~\cite{Cornwall:1981zr,Binosi:2002vk,Binosi:2009qm}
effective charges
in the deep infrared~\cite{Aguilar:2009vn,Aguilar:2010gm}

 \section{Schwinger-Dyson equations for $G(q)$ and $L(q)$}\label{sec:SDE}

Within the standard framework of the SDEs,
the dynamical equations that govern the momentum evolution of 
$G(q)$ and $L(q)$ may be 
deduced directly from 
\1eq{Lambda_GL} by projecting out the 
$g_{\mu\nu}$ and $q_{\mu}q_{\nu}/q^2$ components,  
respectively~\cite{Aguilar:2009nf}; 
the tensorial decomposition~\cite{Aguilar:2018csq}
\be 
H_{\nu\mu}(r,p,q) = g_{\nu\mu}A_1 + r_\mu r_\nu A_2 + q_\mu q_\nu A_3 + r_\mu q_\nu A_4 + r_\nu q_\mu A_5 \,,
\ee
with $A_i := A_i(r,p,q)$, is also employed.

Setting 
$f:= 1- (q\cdot k)^2/( q^2 k^2)$  
and ${\widetilde A}_i :=A_i(-q,k+q,-k)$, 
we obtain
\begin{align}
1 + G(q) =&\, Z_c + \frac{ig^2 C_{\rm A}}{3}\int_k \Delta(k)D(k+q) K_{G} \,,
\nonumber \\
L(q) =&\, \frac{ig^2 C_{\rm A}}{3}\int_k \Delta(k)D(k+q) K_{L} \,,
\label{SDEGL}
\end{align}
where
\begin{align}
K_{G} =& \,
(3 - f) \, 
{\widetilde A}_1 
- (k\cdot q)\, f \,{\widetilde A}_4 
\,, \nonumber \\
K_{L} =&\, 
(3 - 4f) \, {\widetilde A}_1 
- f\left[ 3 q^2 {\widetilde A}_2 + 4(q\cdot k) \, {\widetilde A}_4 \right] \,,
\end{align}
and 
\be
\int_{k} :=\frac{1}{(2\pi)^{4}}\!\int_{-\infty}^{+\infty}\!\!\mathrm{d}^4 k
\label{dqd}
\ee
is the properly regularized measure.  

Note that, due to the 
validity of \1eq{FGL}, 
the $G(q)$ in \1eq{SDEGL}
is renormalized through  
the $Z_c$, namely the ghost renormalization constant.  
For its determination, one must consider 
the corresponding SDE for the ghost dressing function $F(q)$, see, \eg~\cite{Aguilar:2022exk}.
This dynamical equation involves 
the ghost-gluon vertex as one of its ingredients;  
this latter vertex is finite in the Landau gauge, and its 
attendant renormalization constant, $Z_1$, is independent of the cutoff. 
Within the Taylor scheme~\cite{Boucaud:2008gn,Boucaud:2011eh,vonSmekal:2009ae} that we will employ, we have $Z_1 = 1$, 
and the SDE for $F(q)$ assumes the form 
\be
F^{-1}(q) =  Z_c + i g^2 C_{\rm A} \int_k f \Delta(k)D(k+q)   {\widetilde B}_1 \,,
\label{FSDE}
\ee
where ${\widetilde B}_1 := B_1(-q,k+q,-k)$, with $B_1$ 
denoting the classical form factor of the ghost-gluon vertex. 
Then, the expression for $Z_c$ to be used in \1eq{SDEGL} is obtained from 
\1eq{FSDE} by imposing the renormalization condition 
$F(\mu) =1$, where $\mu$ is the renormalization point; throughout this 
work we use $\mu = 4.3$ GeV. 

The numerical treatment of the SDE system formed by \2eqs{SDEGL}{FSDE} proceeds along the lines described in~\cite{Aguilar:2009pp,Aguilar:2009nf}, using up-to-date ingredients 
as external inputs. 
Specifically, we employ general kinematics results for ${\widetilde B}_1$ and ${\widetilde A}_i$, determined from their own SDEs in~\cite{Aguilar:2018csq}; instead,  in~\cite{Aguilar:2009pp,Aguilar:2009nf} these form factors 
were set at their tree-level values. Moreover, we employ directly a fit to the lattice gluon propagator $\Delta(q)$
(lower panel of \fig{fig:G_L}), 
obtained from the same lattice setups used for the determination of $1+G(q)$ and $F(q)$. The resulting SDE solutions 
for $G(q)$ and $L(q)$ will be discussed 
in Sec.~\ref{sec:results}.

\section{Lattice simulation: theory and setup}\label{sec:lattice_setup} 

Both 
the formal definition of 
$G(q)$ in terms of BV fields
and the alternative presented in \1eq{Lambda_GL} are unsuitable for 
performing lattice simulations.
Instead, we will 
take advantage of the 
known equality between 
$G(q)$ and the KO function\footnote{
This function 
is associated  
with a standard confinement criterion~\cite{Kugo:1979gm,Kugo:1995km};
for a review, see~\cite{Alkofer:2000wg} . It is well-known   
that the $u$ simulated on the lattice
does not comply with this criterion
\cite{Nakajima:1999dq,Sternbeck:2006rd}, 
nor do the "decoupling" propagators, see, \eg~\cite{Ilgenfritz:2006he,Bogolubsky:2007ud,Cucchieri:2007md,Oliveira:2012eh} 
and~\cite{Braun:2007bx,Aguilar:2008xm,Fischer:2008uz}.}, to be denoted  
by $u(q)$; specifically, 
in the Landau gauge, we 
have~\mbox{\cite{Kugo:1995km,Grassi:2004yq,Kondo:2009ug,Aguilar:2009pp}}.
\be
G(q) = u(q) \,.
\label{GisU}
\ee
The field-theoretic definition of $u(q)$ that 
has been implemented on the lattice involves
two composite operators,
\begin{equation}
{\cal A}^{a}_{\mu}(x) :=  D_\mu^{a e}(x) \, c^e(x) \,,
\qquad 
{\cal B}^{a}_{\nu}(x) := f^{b c d} A_\nu^d(x) \, \bar{c}^c(x) \,,
\end{equation}
where $D_\mu^{ab}(x) = \partial_\mu \delta^{ac} + g f^{amb} A^m_\mu(x) $ 
is the covariant 
derivative in the adjoint 
representation, 
and 
the two-point function 
$\mathcal{U}^{ab}_{\mu\nu}(q)$, defined 
as (Euclidean space) 
\begin{equation}
 \mathcal{U}^{ab}_{\mu\nu}(q) =  \int d^4 x e^{i q(x-y)}\ \langle 0 | T \left( 
{\cal A}^{a}_{\mu}(x) \,{\cal B}^{b}_{\nu}(y)
    \right)|0\rangle \,,
\label{}
   \end{equation}
where $T$ 
denotes the standard time-ordering operation.
The KO function  
is the scalar co-factor of $\mathcal{U}^{ab}_{\mu\nu}(q)$, 
\begin{equation}
\mathcal{U}^{ab}_{\mu\nu}(q) =
\delta^{a b}\left(\delta^{\mu \nu}-\frac{q_\mu q_\nu}{q}\right) u(q) \,.
\label{kobig}
\end{equation}

An explicit lattice definition for $\mathcal{U}^{ab}_{\mu\nu}(q)$
is given by 
\begin{equation}
\mathcal{U}^{ab}_{\mu\nu}(q)\!=\!\frac{1}{V}\left\langle\sum_{x, y, z}\sum_{c, d, e} e^{-i q \cdot ( x - y )} \left(D_\mu\right)^{a e}\!\!\!(x;z)\left(M^{-1}\right)^{e c}\!\!\!(z;y) f^{b c d} A_{\nu}^d(y) \right\rangle_U
\label{kugolat}
\end{equation}
where $\left\langle \ldots \right\rangle_U$ denotes the Monte-Carlo
sample average over the gauge field configurations $U$, and $M$ is a suitably discretized version of the Faddeev-Popov operator, $\partial_\mu D^{ab}_\mu$; note that the inverse of $M$ is related to the ghost propagator by 
\begin{equation}
 D^{ab}(q) =  \int d^4 x e^{i q(x-y)}\ \langle 0 | T \left(M^{-1}\right)^{a b}\!\!\!(x;y)|0\rangle \,.
\label{ghost_prop}
\end{equation}
With the aid of  \1eq{kobig}, 
the scalar function $u(q)$ is given by
\begin{equation}
u(q) = \frac{1}{24}\sum_{\mu,a}\mathcal{U}^{aa}_{\mu\mu}(q) \,.
\end{equation}

In order to study the KO function on the lattice, we rely on Eq.~(\ref{kugolat}). 
However, for practical reasons, it is convenient to compute $\mathcal{U}^{ab}_{\mu\nu}(q)$ using a point source $y_0$ in the inversion
of the lattice Faddeev-Popov operator, $M$,
\begin{equation}
\mathcal{U}^{ab}_{\mu\nu}(q\!=\!\left\langle\sum_{x, z} \sum_{c, d, e} e^{-i q \cdot ( x - y_{0} )} \left(D_\mu\right)^{a e}\!\!\!(x;z)\left(M^{-1}\right)^{e c}\!\!\!(z;y_{0}) f^{b c d} A_{\nu}^d(y_{0}) \right\rangle_U
\label{U_lat}
\end{equation}
where the lattice definitions for  $D$, $M$, and $A$ can be found in~\mbox{\cite{Sternbeck:2006rd,Silva:2004bv,Suman:1995zg}.
} 

The most important points 
of the lattice simulation 
may be summarized as follows:

($\it i$) The gauge field, $A_\mu^a$, covariant derivative, $D_\mu^{ab}$, and Faddeev-Popov operator, $M_{xy}^{ab}$, are suitably discretized according to~\cite{Sternbeck:2006rd,Brito:2023pqm,Oliveira:2012eh,Cucchieri:2018doy}.

($\it ii$) The Landau gauge is fixed by identifying the configurations belonging to the Gribov region; this is achieved  through the maximization of a certain functional, whose discretized form is given in ~\cite{Sternbeck:2006rd,Brito:2023pqm}. Numerically, the required maximization is performed with the Fourier Accelerated Steepest Descent method~\cite{Davies:1987vs}.

($\it iii$) Since $M_{xy}^{ab}$ is a singular operator, it can only be inverted in the subspace orthogonal to its zero modes. Nevertheless, this is sufficient for computing $u(q)$ for $q\neq 0$~\cite{Suman:1995zg}. In this case, the inversion problem in the orthogonal subspace can be transformed into a sparse linear system of equations, as explained in detail in~\cite{Sternbeck:2006rd,Brito:2023pqm,Suman:1995zg}. Then, since $M_{xy}^{ab}$ is real and symmetric, the resulting system can be solved efficiently with the conjugate gradient method~\cite{doi:10.1137/1.9781611971538}.

($\it iv$) 
We consider quenched lattice ensembles generated with the Wilson gauge action~\cite{Wilson:1974sk}, with  $\beta=6.0$. In this case, the Sommer parameter method~\cite{Necco:2001xg} yields a lattice spacing of $a = 0.0962$~fm ($a^{-1} = 2.05$~GeV)~\cite{Boucaud:2018xup}. We employ two setups, with lattice volumes $V = 64^4$, and $V=80^4$, in lattice units; the corresponding physical volumes extend from $~(6\textrm{ fm})^4$ to $~(8\textrm{ fm})^4$. Further details on sampling, gauge
fixing, and definitions can be found in~\cite{Duarte:2016iko}.

($\it v$) The number of gauge field configurations considered for the present study was $700$ for $V=64^4$, and $500$ for $V=80^4$.

($\it vi$) The computer simulations were performed with the help of the Chroma library~\cite{Edwards:2004sx},  using PFFT~\cite{Pippig:2013} for the necessary Fourier transforms.

\section{Results}\label{sec:results} 

Even though the lattice simulation 
uses the expression in \1eq{kobig}
as its starting point, in 
what follows we present the results 
in terms of $G(q)$, which is the 
focal point of this study, rather than $u(q)$. 
Evidently, by virtue of \1eq{GisU}, 
the results may be recast trivially in terms of  $u(q)$, 
by setting \mbox{$G \!\to\!u$} throughout.

\begin{figure}[t]
  \begin{center}
    \includegraphics[width=0.4\textwidth]{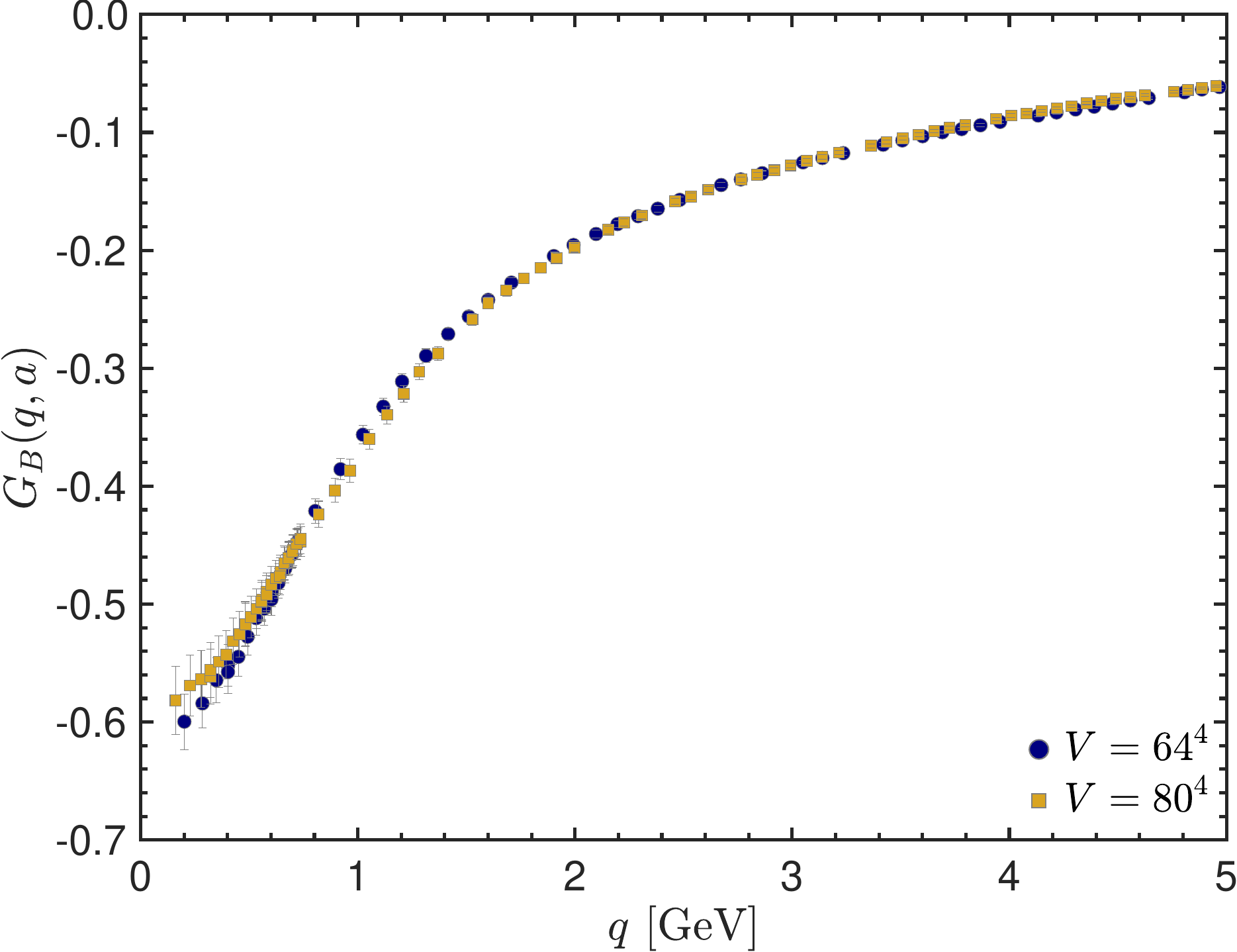}
  \end{center}
 \caption{Bare lattice results 
 for the BV function, 
 $G_{\srm B}(q,a)$. }
\label{fig:G_bare}
\end{figure}

Let us denote the bare $G(q)$ obtained on the lattice by $G_{\srm B}(q,a)$, making manifest the dependence on the lattice spacing $a$. The result for $G_{\srm B}(q,a)$ from our simulations are shown in \fig{fig:G_bare}, for both lattice volumes. We note that the 
volume dependence is rather weak; indeed, the data from both volumes agree within errors.

Ideally, 
in order to obtain a reliable 
comparison with the SDE results, the $G_{\srm B}(q,a)$ should be extrapolated to the continuum limit, $a \to 0$, and be renormalized using the Taylor scheme. 
Instead, in what follows 
we adopt a slightly modified version of the procedure developed in~\cite{Sternbeck:2006rd}, whereby  $G_{\srm B}(q,a)$ is renormalized by imposing \1eq{FGL}.

To begin with, the continuum limit 
\mbox{$G_{\srm B}(q) := G_{\srm B}(q,0)$} is related to the lattice $G_{\srm B}(q,a)$ by
\be 
G_{\srm B}(q,a) = G_{\srm B}(q) + s(q) \,,
\ee
where $s(q)$ accounts for lattice artifacts, and, in principle, depends on the momentum $q$. However, the momentum dependence of $s(q)$ cannot be determined by the analysis employed below. Therefore, 
in order to proceed, 
we {\it assume} that the dependence of $s(q)$ on the momentum is fairly mild~\cite{Oliveira:2012eh} and treat it as a constant, \ie $s(q)\to s$.

Then, the renormalized lattice $1 + G(q)$ in the Taylor scheme is given by
\be 
1 + G(q) = Z_c[1 + G_{\srm B}(q,a) - s] \,, \label{G_lat_ren}
\ee
where the parameter $s$ acts as an effective subtractive renormalization constant. We emphasize  that subtractive terms in the renormalization have been shown to improve the agreement with the continuum theory in studies of the quark propagator and the quark-gluon vertex~\cite{Oliveira:2018lln,Kizilersu:2021jen}.

The value of $Z_c$ can be determined formally by imposing a renormalization condition, \ie a value for the renormalized \mbox{$1 + G(\mu)$}, at some scale $\mu$. However, since in the Taylor scheme $F(\mu)=1$, and $L(\mu)$ generally does not vanish, \1eq{FGL} implies that we cannot impose $1 + G(\mu) = 1$. 

Instead, we determine the value of $1 + G(\mu)$ by first computing $1 + G(q)$ at one loop,
\be 
1 + G^{(1)}(q) = 1 + \frac{\alpha_s C_{\rm A}}{16\pi}\left[ 3 \ln(q^2/\mu^2) - 2\right] \,,
\ee
with $\alpha_s = g^2/4\pi$.
Then, using the value of $\alpha_s = 0.216$ for the Taylor coupling at $\mu = 4.3$~GeV~\cite{Boucaud:2008gn} we obtain
\be 
1 + G^{(1)}(\mu) = 0.974 \,. \label{Gmu_pert}
\ee
Interestingly, the SDE result yields $1 + G(\mu) = 0.973$, indicating that the perturbative regime of $1 + G(q)$ has been 
safely reached for \mbox{$q = \mu = 4.3$~GeV}. Hence, we can fix the scale of the lattice $1 + G(q)$ by imposing $1 + G(\mu) = 1 + G^{(1)}(\mu)$, such that
\be 
Z_c = \frac{1 +  G^{(1)}(\mu) }{1+G_{\srm B}(\mu,a) - s} = \frac{0.974}{0.922 - s} \,, \label{G_lat_Zc}
\ee
where used $G_{\srm B}(\mu,a) = -0.078$, obtained from the bare data of \fig{fig:G_bare}.

At this point, $Z_c$ still depends on the unknown constant $s$. To fix its value, we impose the validity of 
\1eq{FGL}, \ie 
\be 
Z_c\left[ 1 + G_{\srm B}(q,a) - s \right] + L(q) = F_{\!\!\srm{lat}}^{-1}(q) \,, \label{FGL_lat}
\ee
where $F_{\!\!\srm{lat}}(q)$ is the renormalized lattice ghost dressing function, obtained self-consistently from the same lattice setups used to compute $G_{\srm B}(q,a)$. 

Now, the function $L(q)$ is not directly computed on the lattice. Nevertheless, for small $q$ we can assume for $L(q)$ a Taylor expansion,
\be 
L(q) = a_1 q^2 + a_2 q^4 + {\cal O}(q^6) \,, \label{L_expansion}
\ee
whose $0$-th order term vanishes on account of $L(0) = 0$.

Then, we can determine the parameters $s$, $a_1$, and $a_2$, by rephrasing \1eq{FGL_lat} as a $\chi^2$ minimization problem. Specifically, we minimize the $\chi^2$ defined by
\be 
\chi^2 := \sum_{i}\left\lbrace Z_c\left[ 1 + G_{\srm B}(q_i,a) - s \right] + a_1 q_i^2 + a_2 q_i^4 - F_{\!\!\srm{lat}}^{-1}(q_i) \right\rbrace^2 \,, \label{chi}
\ee
where $q_i$ are the lattice points, and $Z_c$ is given by \1eq{G_lat_Zc}.

\begin{figure}[t]
  \centering
  \includegraphics[width=0.4\textwidth]{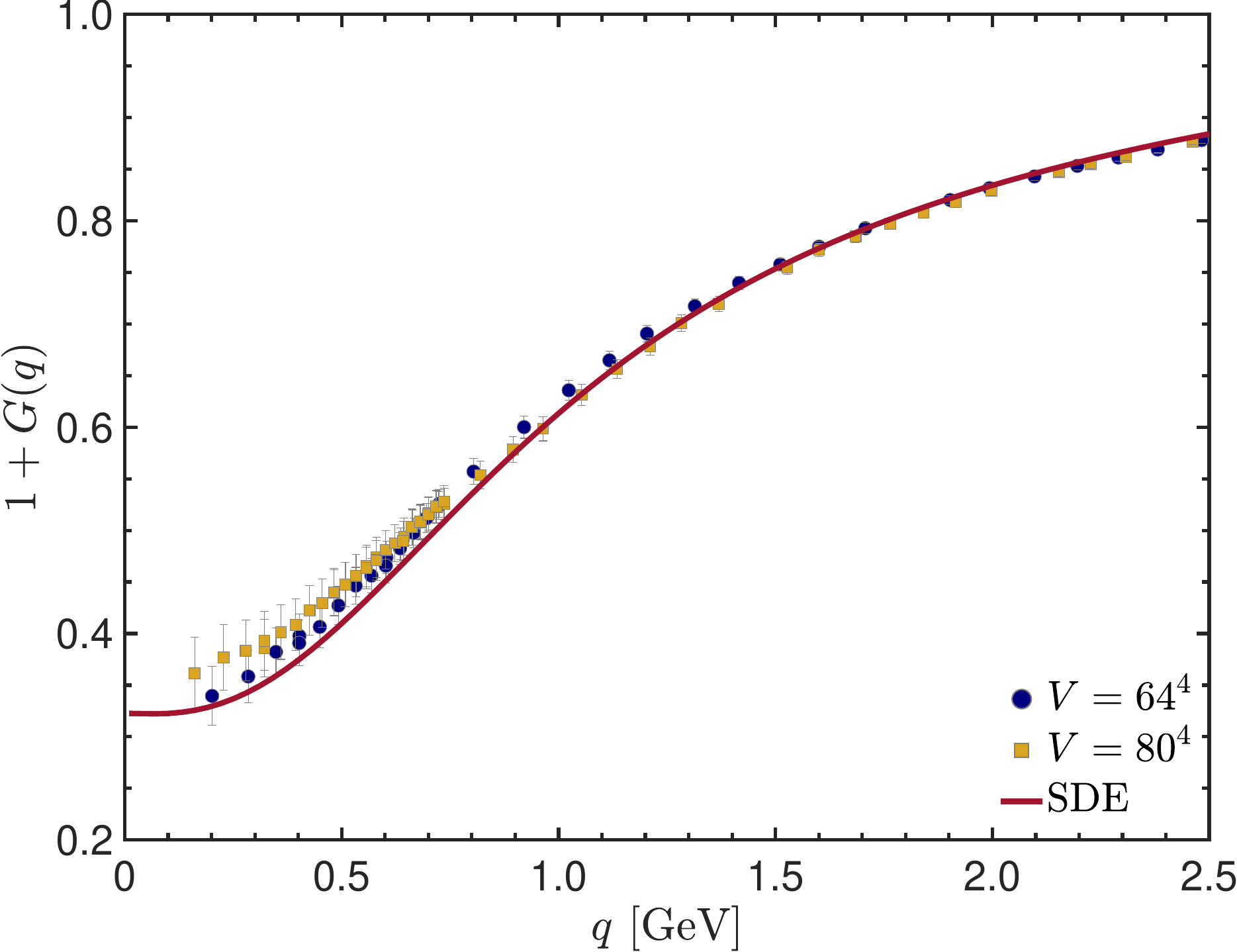}\\
  \includegraphics[width=0.4\textwidth]{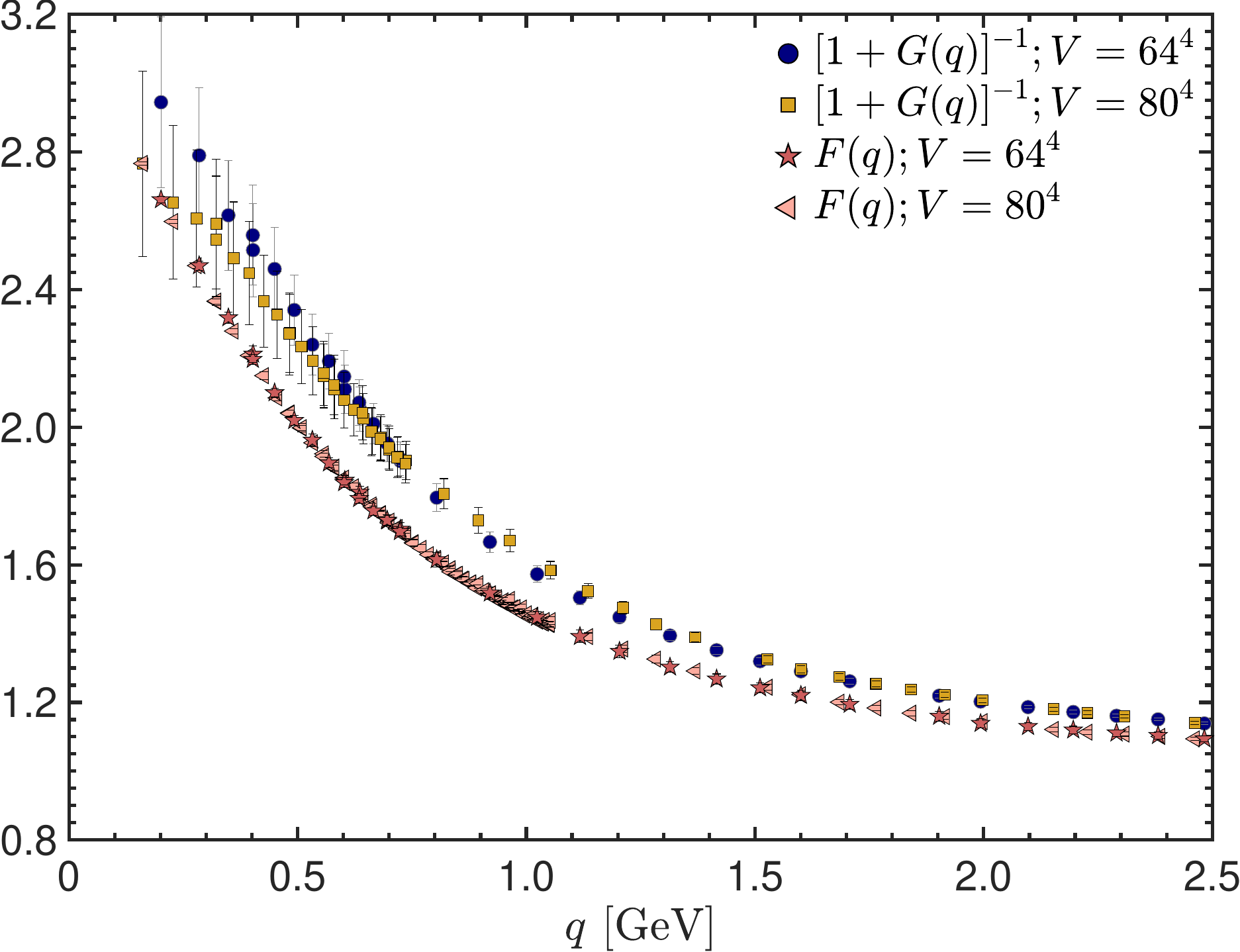}
\caption{ Top: Results for $1 + G(q)$  obtained from the lattice (points) and the SDE prediction (red curve). Bottom: Comparison of $[1+G(q)]^{-1}$ (circles/squares) and $F(q)$ (stars/triangles). }
\label{fig:G_F}
\end{figure}

Since \1eq{chi} is the result of a Taylor expansion, it is supposed to be used only for small $q$; specifically, we choose the points $q_i < 0.5$~GeV. Consequently, there are too few points to reliably determine the parameters $s$, $a_1$ and $a_2$ for the two lattice setups individually. Instead, we apply the above procedure to the \emph{combined} data points of both volumes, for which there are $18$ points in the fitting window. We have verified that for the combined data, the value of $s$ is stable against varying the order of the Taylor expansion in \1eq{L_expansion}, as well as the fitting window. Specifically, we obtain $s = 0.121$, $a_1 = 0.458~\text{GeV}^{-2}$, and $a_2 = - 0.905~\text{GeV}^{-4}$, for which $\chi^2 = 2\times10^{-3}$. 

Using the above value of $s$ in \2eqs{G_lat_ren}{G_lat_Zc},  we obtain the renormalized lattice $1 + G(q)$, shown as points on the top panel of \fig{fig:G_F}. On the same panel we show also the SDE prediction (red continuous), finding an excellent agreement. On the bottom panel of \fig{fig:G_F}, we compare $F(q)$ to $[1+G(q)]^{-1}$; evidently, 
\1eq{FG0} is satisfied within errors, as a consequence of the renormalization procedure employed.

\begin{figure}[t]
  \centering
  \includegraphics[width=0.4\textwidth]{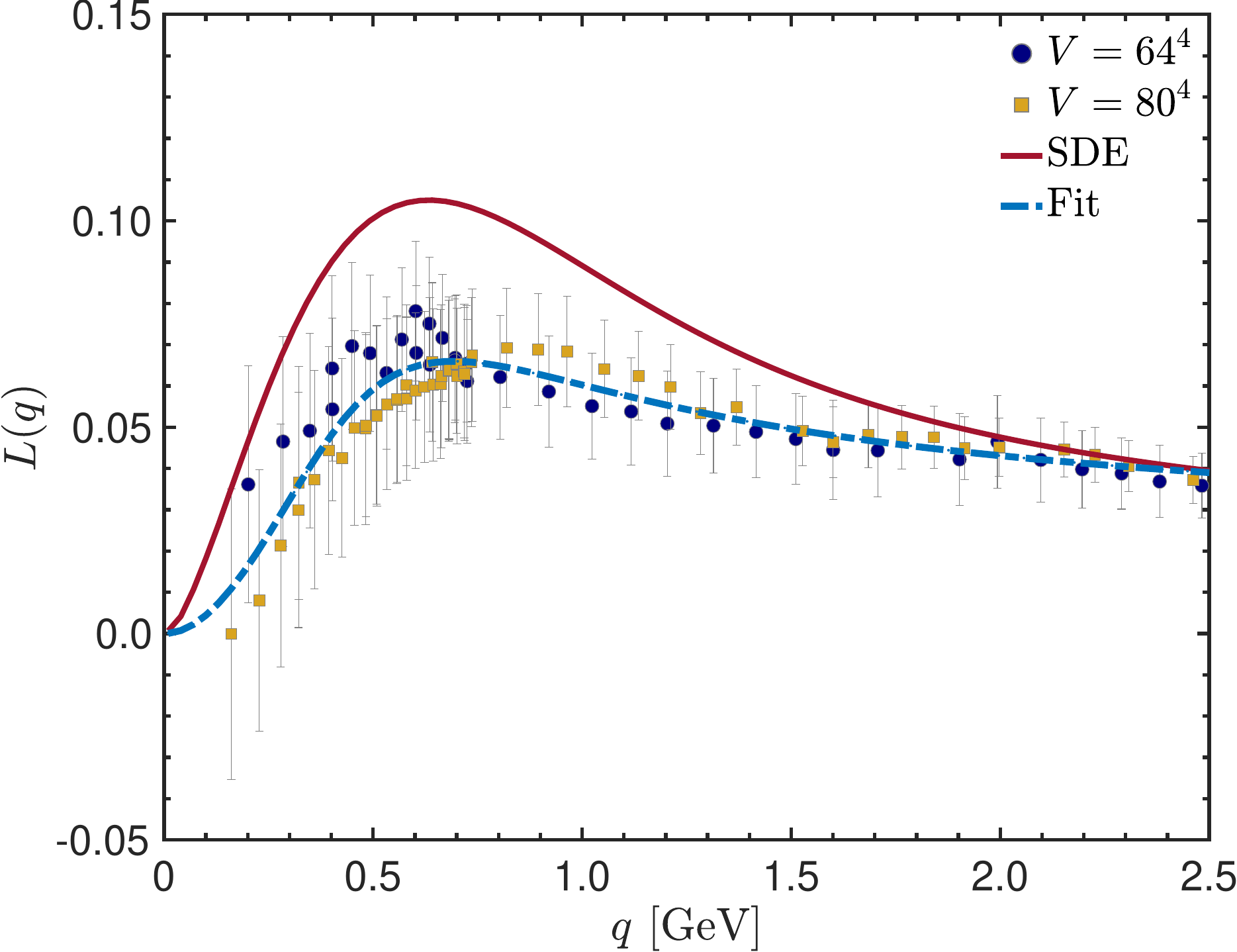}\\
  \includegraphics[width=0.4\textwidth]{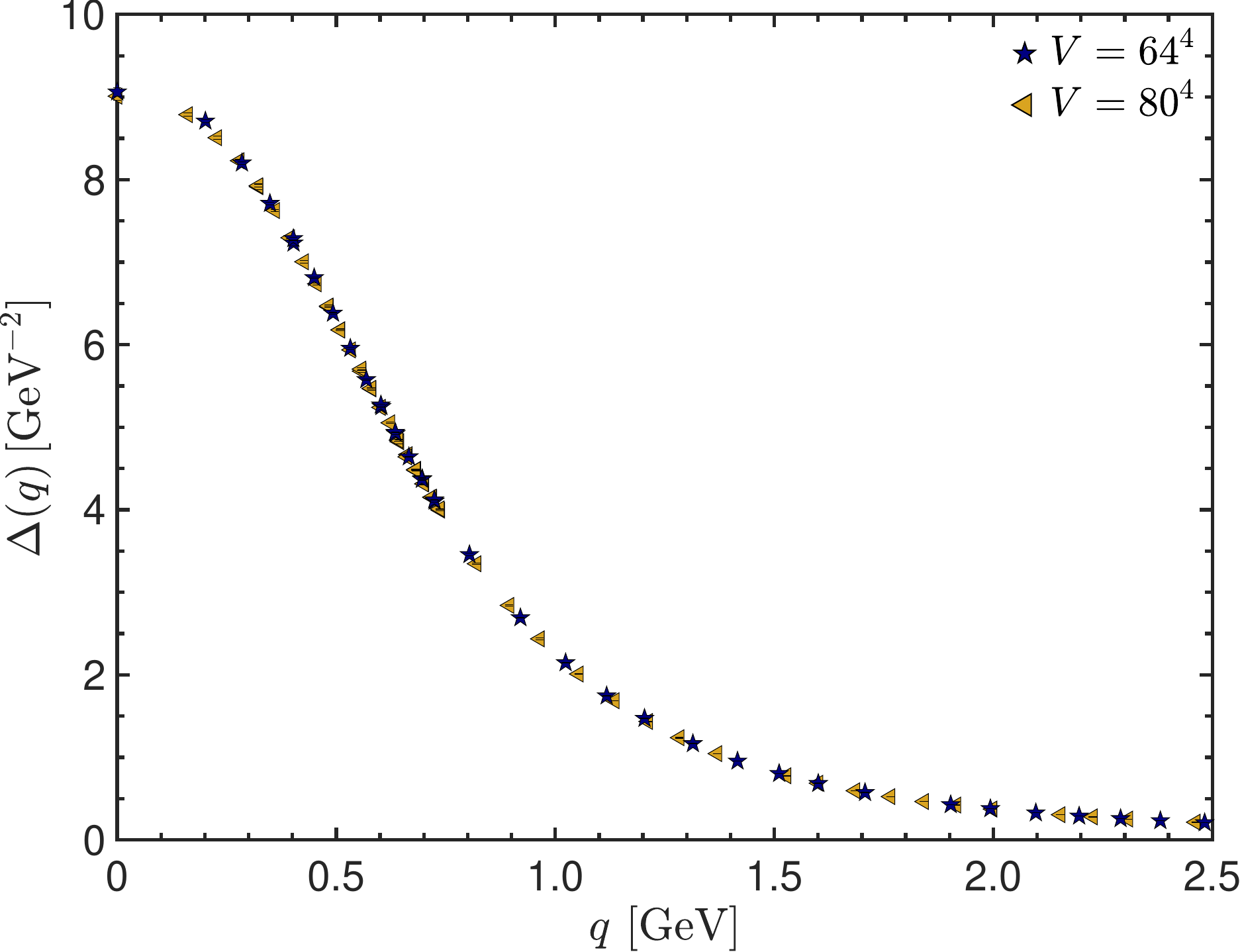}  
\caption{ Upper panel: $L(q)$ obtained from the lattice $1+G(q)$ and $F(q)$ of \fig{fig:G_F} (points) compared to the SDE prediction (red continuous). Also shown is a fit given by \1eq{L_fit} (blue dashed). Lower panel: The gluon propagator simulated on the same lattices as the BV function~\cite{Dudal:2018cli}.}
\label{fig:G_L}
\end{figure}

The comparison between 
lattice and  
SDE results may be taken a step further, by considering 
the full momentum dependence of the 
form factor $L(q)$. 
In 
particular, a lattice-derived result for $L(q)$ may be obtained 
from \1eq{FGL}, by 
substituting in it the 
lattice data for 
$F(q)$ and $1 + G(q)$. 
The outcome of this procedure 
is shown in \fig{fig:G_L};  
the comparison with the  
SDE prediction (red continuous curve) reveals a reasonable agreement. 
The blue curve represents a fit to the lattice $L(q)$, given by the 
simple functional form
\be 
L(q) = \frac{ q^2/c_1^2 + (q^2/c_2^2)^2 }{ 1 + q^2/t_1^2 + (q^2/t_2^2)^2\ln^{d_{\srm L}}( 1 + q^2/\Lambda_{\srm T}^2)  } \,, \label{L_fit}
\ee
where $\Lambda_{\srm T} = 425$~MeV is the value of $\Lambda_{\srm{QCD}}$ in the Taylor scheme~\cite{Boucaud:2008gn}, and $d_{\srm L} = 35/44$ is the anomalous dimension of $L(q)$~\cite{Binosi:2016xxu}. The fitting parameters are given by $c_1^2 = 2.18~\text{GeV}^2$, $c_2^2 = 0.575~\text{GeV}^2$, $t_1^2 = 0.0981~\text{GeV}^2$, and $t_2^2 = 0.188~\text{GeV}^2$. Note that \1eq{L_fit} satisfies $L(0) = 0$ by construction.

\section{RGI running interaction strength}\label{sec:dhat}


As has been explained in a series of articles~\cite{Binosi:2009qm,Aguilar:2009nf,Binosi:2014aea}, the fundamental relation of \1eq{Zg_ZA} allows for the definition of a propagator-like 
RGI quantity, exactly as happens 
in QED when the photon 
vacuum polarization is multiplied by $e^2$. 
In particular, since $\wh\Delta(q)$ renormalizes 
through the $Z_B$ introduced in 
\1eq{Zg_ZA}, \ie
\be
\wh\Delta_{{\rm \s{R}}}(q) = Z_{B}^{-1} \wh\Delta(q) \,,
\ee
the combination 
\be 
{\widehat d}(q) := \alpha_s{\wh \Delta(q)} \,, \label{dhat}
\ee 
is RGI by virtue of \1eq{Zg_ZA}. Indeed, ${\widehat d}(q)$
retains exactly the same form 
before and after renormalization, and, consequently, does 
not depend on the renormalization point $\mu$, nor on the renormalization 
scheme employed. 

The BV function enters into the 
definition of ${\widehat d}(q)$ 
when the central relation in  \1eq{propBQI} is invoked, 
\be 
{\widehat d}(q) :=  \frac{\alpha_s\Delta(q)}{[1 + G(q)]^2} \,; \label{dhatG}
\ee 
in this form, ${\widehat d}(q)$ is 
known in the literature as the ``RGI running interaction strength''~\cite{Binosi:2014aea}.

%
\begin{figure}[!t]
  \centering
 \includegraphics[width=0.75\linewidth]{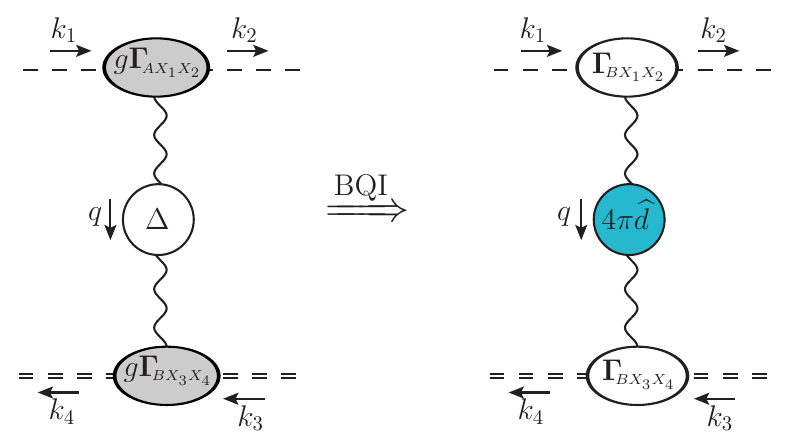}\hfil 
\caption{Diagrammatic representation of the BQI-induced rearrangement in the case of a generic
$X_1 X_2\to X_3 X_4$  scattering, corresponding to the S-matrix element ${\cal T}_{X_1 X_2\to X_3 X_4}$
of \1eq{qq_scat}.}
\label{fig:scattering}
\end{figure}

What is particularly interesting 
about ${\widehat d}(q)$
is that it does not represent a mere 
field-theoretic construct, but 
admits a clear physical interpretation. 
Specifically, 
${\widehat d}(q)$ 
constitutes a common component of any two-to-two physical processes,
\mbox{$X_1(k_1) X_2(k_2)\to X_3(k_3) X_4(k_4)$}, 
where an off-shell gluon, 
carrying momentum 
\mbox{$q = k_1-k_2=k_3-k_4$}, is exchanged, see \fig{fig:scattering}.
In particular, with the 
aid of the BQIs in 
\2eqs{propBQI}{quark_BQI}, 
the S-matrix element
${\cal T}_{X_1 X_2\to X_3 X_4}$
may be cast in the form 
\begin{align} 
{\cal T}_{X_1 X_2\to X_3 X_4} =& \,\, \left\lbrace g \,{\mathbf\Gamma}_{\!A  X_1 X_2} \right\rbrace \Delta(q) \left\lbrace g \,{\mathbf \Gamma}_{A X_3 X_4} \right\rbrace \nonumber\\
=& \,\, \left\lbrace g \, [1 + G(q)]^{-1} {{\mathbf\Gamma}}_{\!B X_1 X_2}
\right\rbrace \Delta(q) \left\lbrace g \, [1 + G(q)]^{-1}{ {\mathbf \Gamma}}_{\! B X_3 X_4} \right\rbrace \nonumber\\
=& \,\,  {{\mathbf\Gamma}}_{\!B X_1 X_2} 
\underbrace{\left\lbrace g^2 [1 + G(q)]^{-2} \Delta(q)  \right\rbrace}_{4\pi \, {\widehat d}(q)}{{\mathbf \Gamma}}_{\! B X_3 X_4} 
\,.
\label{qq_scat}
\end{align}
Due to the general validity of the 
BQIs, the steps leading to the 
last line of \1eq{qq_scat}
may be followed regardless of the 
detailed nature of the 
initial and final states. 
In that sense, ${\widehat d}(q)$ 
captures the process-independent 
contribution, common to all 
such processes, whilst the 
process-dependence,  
\ie the part that carries 
the information about the specific nature 
of the initial and final states,  
is encoded in the vertices 
${{\mathbf\Gamma}}_{\!B X_1 X_2}$
and ${{\mathbf \Gamma}}_{\! B X_3 X_4}$. 

The quantity ${\widehat d}(q)$
has mass dimension of $-2$; from it 
one may define the dimensionless RGI interaction 
${\cal I}(q)$~\cite{Binosi:2014aea} 
\be 
{\cal I}(q) := q^2 {\widehat d}(q) \,. 
\label{Ical}
\ee
As explained in~\cite{Binosi:2014aea}, this quantity provides the strength required in order to describe ground-state hadron observables using SDEs in the matter sector of the theory. 
As was argued there, the physics encoded in ${\cal I}(q^2)$  
reconciles nonperturbative continuum QCD with   
ab initio predictions of basic hadron properties.

\begin{figure}[t]
  \centering
  \includegraphics[width=0.4\textwidth]{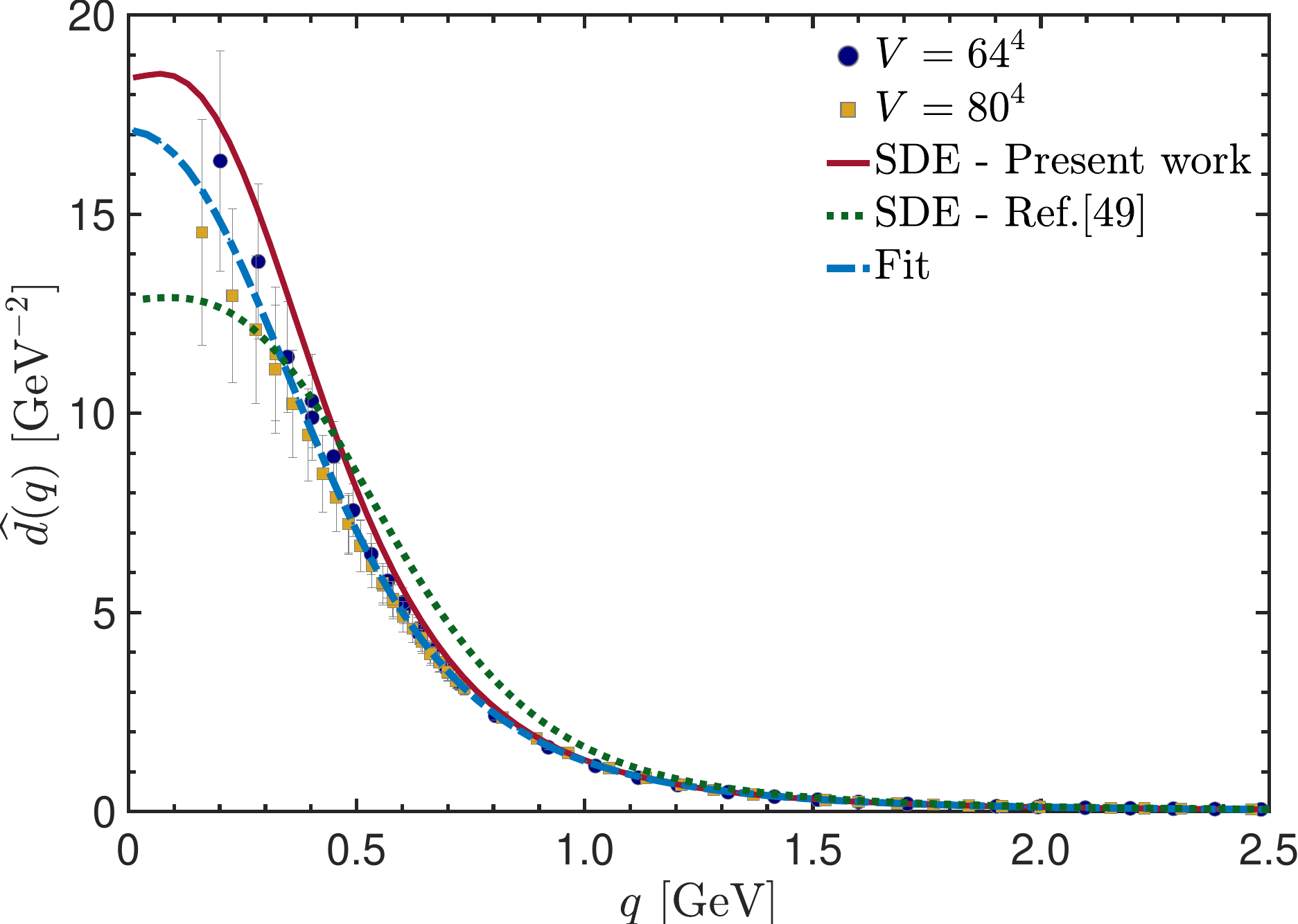}\\
  \includegraphics[width=0.4\textwidth]{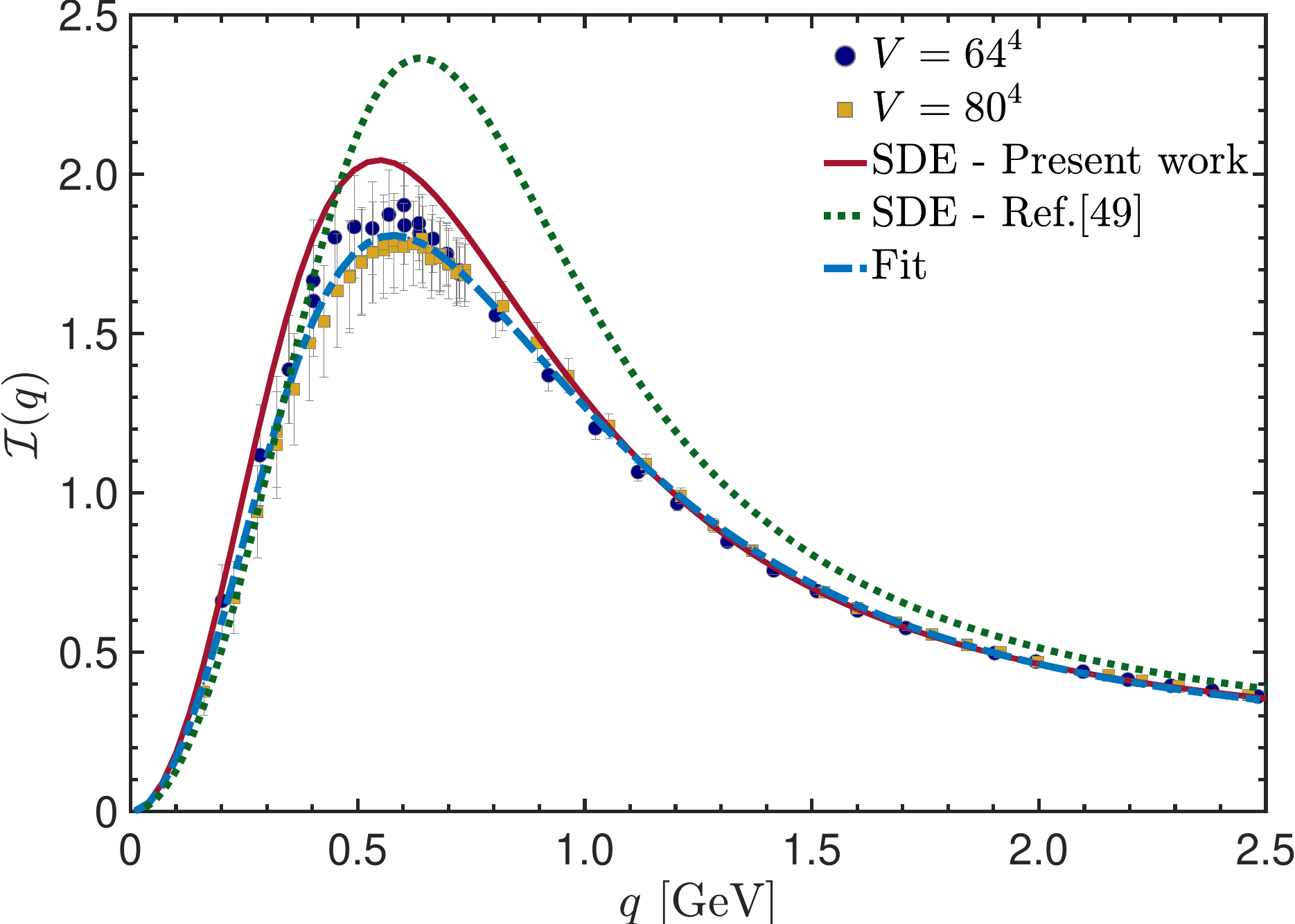}
\caption{ Lattice results for ${\widehat d}(q)$ (top) and ${\cal I}(q)$ (bottom), represented by points. On both panels,  the present SDE result  (red continuous curve) and that of~\cite{Binosi:2014aea} (green dotted)  are shown for comparison. The fit to the lattice data, given by \1eq{dhat_fit} is shown in blue dashed.}
\label{fig:dhat_I}
\end{figure}

In \fig{fig:dhat_I},  
the red continuous curves correspond to 
the ${\widehat d}(q)$ (top) and ${\cal I}(q)$ (bottom), obtained by combining the $1+G(q)$ from \fig{fig:G_F} with the gluon propagator, self-consistently simulated on the same lattice setups (see lower panel 
of \fig{fig:G_L}). For the Taylor coupling at $\mu = 4.3$~GeV, we use $\alpha_s = 0.216$~\cite{Boucaud:2008gn};
note that, due to the RGI nature of ${\widehat d}(q)$, any other 
set $\{\mu, \alpha_s(\mu)\}$
yields precisely the same 
answer. 
In addition, the 
${\widehat d}(q)$ and ${\cal I}(q)$ obtained 
from the SDE analysis of~\cite{Binosi:2014aea} 
are displayed as green dotted curves. Note that 
the present treatment 
leads to closer agreement between 
lattice and SDE results; we have 
confirmed that this is mainly due 
to the non-perturbative dressing of the 
form factors ${\widetilde B}_1$ and ${\widetilde A}_i$,  discussed in the  
last paragraph of Sec.~\ref{sec:SDE}.

Moreover, in \fig{fig:dhat_I} we show as a blue dashed curve a fit to the lattice ${\widehat d}(q)$, given by the functional form
\begin{align} 
{\widehat d}(q) =&\, \frac{ {\widehat d}(0)\left[1 + a_1 q^2 + a_2 q^4 \right] }{1 + b_1 q^2 + b_2 q^4 + \beta_0 {\widehat d}(0) a_2 q^6\ln(1+q^2/\Lambda_{\srm T}^2) } \,, \label{dhat_fit}
\end{align}
where $\beta_0 = 11/(4\pi)$ is the first coefficient of the QCD beta function, $\Lambda_{\srm T} = 425$~MeV~\cite{Boucaud:2008gn}, and the fitting parameters are ${\widehat d}(0) = 17.1~\text{GeV}^{-2}$, $a_1 = 0.0585~\text{GeV}^{-2}$, \mbox{$a_2 = 0.0274~\text{GeV}^{-4}$}, $b_1 = 3.51~\text{GeV}^{-2}$, $b_2 = 9.33~\text{GeV}^{-4}$. Note that \1eq{dhat_fit} enforces the one loop-running of the QCD coupling in the ultraviolet, \ie ${\cal I}(q)\to [\beta_0\ln(q^2/\Lambda_{\srm T}^2)]^{-1}$ at large $q$.

\section{Conclusions}
The quantitative agreement 
between the BV function simulated on the 
lattice and the corresponding 
SDE results exposes once again the  
underlying consistency of a large number of 
concepts and techniques, developed over a period of several years, 
see, \eg~\cite{Ferreira:2023fva}. Especially 
interesting in that regard is the interplay between the ghost and gauge sectors of the theory, 
which are nontrivially intertwined by the 
SDEs. Note, in particular, that the lattice 
gluon propagator in \fig{fig:G_L} is used as input in the SDEs of \2eqs{SDEGL}{FSDE},
which produce the prediction for $G(q)$ that is  subsequently compared with the 
corresponding
lattice result in  \fig{fig:G_F}. Let us 
finally emphasize that 
the effective interactions ${\widehat d}(q)$ 
and ${\cal I}(q)$
shown in \fig{fig:dhat_I} correspond to the 
pure Yang-Mills case; for phenomenological 
applications they must be modified  
to include effects from dynamical quarks, 
in the spirit of~\cite{Binosi:2016xxu,Cui:2019dwv,Aguilar:2019uob}. In fact, 
the results of the present 
work bolster up our confidence in the 
SDE derivations that lead to the 
"unquenching" of these RGI quantities. 

\section*{Acknowledgements}
The authors acknowledge the computing time provided by the Laboratory for Advanced Computing at the University of Coimbra 
(FCT contracts  2021.09759.CPCA and  2022.15892.CPCA.A2). Work supported by FCT contracts CERN/FIS-PAR/0023/2021, \href{https://doi.org/10.54499/UIDB/04564/2020}{UIDB/04564/\-2020}, and  \href{https://doi.org/10.54499/UIDP/04564/2020}{UIDP/04564/2020}. P.~J.~S.  acknowledges
financial support from FCT  under Contract No.~\href{https://doi.org/10.54499/CEECIND/00488/2017/CP1460/CT0030}{CEECIND/00488/2017}. A.~C.~A is supported by the CNPq grant \mbox{310763/2023-1}, and acknowledges financial support from project 464898/2014-5 (INCT-FNA). M.~N.~F. and J.~P. are supported by the Spanish MICINN grant PID2020-113334GB-I00. M.~N.~F. acknowledges financial support from Generalitat Valenciana through contract \mbox{CIAPOS/2021/74}. J.~P. also acknowledges funding from the Generalitat Valenciana grant CIPROM/2022/66.
This work was granted access to the HPC resources of the
PDC Center for High Performance Computing at the KTH
Royal Institute of Technology, Sweden, made available
within the Distributed European Computing Initiative by
the PRACE-2IP, receiving funding from the European
Community’s Seventh Framework Programme (FP7/
2007-2013) under grant agreement No. RI-283493. The
use of Lindgren has been provided under DECI-9 project
COIMBRALATT. We acknowledge that the results of this
research have been achieved using the PRACE-3IP project
(FP7 RI- 312763) resource Sisu based in Finland at CSC.
The use of Sisu has been provided under DECI-12 project
COIMBRALATT2.



\end{document}